\documentclass[12pt,a4paper]{article}
\pdfoutput=1
\usepackage{color}
\usepackage{amssymb,amsmath,bm,bbm}
\usepackage{epsf}
\usepackage{epsfig}
\usepackage{afterpage}
\usepackage{longtable}
\usepackage[dvipsnames]{xcolor}
\usepackage[linktoc=page,bookmarks=false,colorlinks=false,
linkbordercolor=RoyalBlue,citebordercolor=ForestGreen,urlbordercolor=CornflowerBlue]{hyperref}
\usepackage{latexsym,mathrsfs,dsfont}
\usepackage[normalem]{ulem} 
\usepackage[compress]{cite}
\usepackage{graphicx}
\usepackage{url}
\usepackage{paralist}
\usepackage{bbold}

\newcommand{\f}{\frac}
\newcommand{\ord}{\mathcal{O}}
\def\as{\alpha_s}
\def\aspi{\frac{\as}{4\pi}}

\newcommand{\gev}{\, {\rm GeV}}
\newcommand{\mev}{\, {\rm MeV}}

\newcommand{\bsi}{B_6^{(1/2)}}
\newcommand{\bei}{B_8^{(3/2)}}

\def\epe{\varepsilon'/\varepsilon}
\newcommand{\beq}{\begin{equation}}
\newcommand{\eeq}{\end{equation}}
\newcommand{\be}{\begin{equation}}
\newcommand{\ee}{\end{equation}}
\newcommand{\bi}{\begin{itemize}}
\newcommand{\ei}{\end{itemize}}
\newcommand{\ba}{\begin{array}}
\newcommand{\ea}{\end{array}}
\newcommand{\beqa}{\begin{eqnarray}}
\newcommand{\eeqa}{\end{eqnarray}}
\newcommand{\bea}{\begin{eqnarray}}
\newcommand{\eea}{\end{eqnarray}}
\newcommand{\beqn}{\begin{eqnarray}}
\newcommand{\eeqn}{\end{eqnarray}}

\newcommand{\nn}{\nonumber}

\definecolor{red}{cmyk}{0,1,1,0.4}

\usepackage{fancyhdr}
\advance \headheight by 15.0truept       

\begin{document}


\vspace{-14mm}
\begin{flushright}
        {AJB-18-2}\\
CP3-18-25
\end{flushright}

\vspace{8mm}

\begin{center}
{\Large\bf
\boldmath{Dual QCD Insight into  BSM Hadronic Matrix Elements for $K^0-\bar K^0$ Mixing from Lattice QCD   }}
\\[12mm]
{\bf \large  Andrzej~J.~Buras ${}^a$ and Jean-Marc G\'erard${}^b$ \\[0.8cm]}
{\small
${}^a$TUM Institute for Advanced Study, Lichtenbergstr.~2a, D-85748 Garching, Germany\\
Physik Department, TU M\"unchen, James-Franck-Stra{\ss}e, D-85748 Garching, Germany\\[2mm]
${}^b$ Centre for Cosmology,
Particle Physics and Phenomenology (CP3), Universit{\'e} catholique de Louvain,
Chemin du Cyclotron 2,
B-1348 Louvain-la-Neuve, Belgium}
\end{center}

\vspace{8mm}

\abstract{%
\noindent
We calculate BSM hadronic matrix elements  for  $K^0-\bar K^0$ mixing in the Dual QCD approach (DQCD). 
The ETM, SWME and RBC-UKQCD lattice collaborations find  the matrix elements of the BSM density-density operators  $\mathcal{O}_i$ with
$i=2-5$ to be rather different from their vacuum insertion values (VIA) with 
$B_2\approx 0.5$, $B_3\approx B_5\approx 0.7$ and $B_4\approx 0.9$ 
at $\mu=3\gev$ to be compared with $B_i=1$ in the VIA.
We demonstrate that this pattern can 
be reconstructed within the DQCD through the non-perturbative 
meson evolution from very low scales, where factorization of matrix elements is valid, to scales $\ord(1\gev)$ with subsequent perturbative quark-gluon evolution to $\mu=3\gev$. This turns out to be possible in spite of a very different pattern displayed at  low scales with  $B_2=1.2$, $B_3=3.0$, $B_4=1.0$ and $B_5\approx 0.2$ in  the large $N$ limit, $N$ being the number of colours. Our 
results imply that the inclusion of 
 meson evolution in the phenomenology of any non-leptonic transition like $K^0-\bar K^0$ mixing  and $K\to\pi\pi$ decays is mandatory. While meson evolution, as demonstrated in our paper, is hidden in LQCD results, to our knowledge DQCD is the only analytic approach for non-leptonic transitions and decays which takes this important QCD dynamics into account.}

\setcounter{page}{0}
\thispagestyle{empty}
\newpage

\tableofcontents

\section{Introduction}
The $K^0-\bar K^0$  mixing induced CP-violation in $K\to\pi\pi$ decays described by the parameter $\varepsilon_K$ and  the $K_L-K_S$ mass difference $\Delta M_K$  have played  very important role in the construction of the Standard 
Model (SM) and more recently in tests of its possible extensions. For 
recent reviews see \cite{Buras:2013ooa,Buras:2016qia}. In spite 
of modest tensions in $\varepsilon_K$ \cite{Blanke:2016bhf}, in particular in correlation with mass differences 
$\Delta M_{d,s}$, the SM describes the data on $\varepsilon_K$ reasonably well.
The situation with $\Delta M_K$ is unclear at present because its value within the SM is subject to large long distance (LD) \cite{Bijnens:1990mz,Bai:2014cva,Christ:2015pwa,Buras:2014maa} and short distance (SD) \cite{Brod:2011ty} uncertainties. Consequently 
there is still considerable room for new physics (NP) contributions in $\varepsilon_K$ and $\Delta M_K$ which 
are generally governed by new local operators $\mathcal{O}_i$ with $i=2-5$ 
beyond the single SM operator $\mathcal{O}_1$. These BSM density-density operators are known 
to have enhanced hadronic matrix elements implying very strong constraints on various
NP models.

While the Wilson coefficients of the full set of $\Delta S=2$ operators  $\mathcal{O}_i$ 
have been calculated at the NLO level long time ago  \cite{Buras:1990fn,Ciuchini:1997bw,Buras:2000if}, only in this decade have their hadronic matrix 
elements been calculated in QCD with respectable precision. In the case of the SM 
operator  $\mathcal{O}_1$ represented by the parameter $\hat B_K$ several 
lattice QCD collaborations confirmed with high precision its value 
 in the framework of Dual QCD approach (DQCD) \cite{Bardeen:1987vg,Buras:2014maa} finding, in agreement with the latter, that its value is rather different
from its  vacuum insertion estimate (VIA) and very close to its large $N$ limit
$\hat B_K=3/4$. While lattice QCD was not able to explain this result, such an
explanation is provided by DQCD as demonstrated in \cite{Buras:2014maa} 
where, relative to the first paper \cite{Bardeen:1987vg},
vector meson contributions have been included and the  matching 
to short distance contributions thereby improved. 

Recently very useful results on hadronic matrix elements of  the BSM operators $\mathcal{O}_i$ with $i=2-5$ have been obtained  by 
the ETM, SWME and RBC-UKQCD lattice collaborations \cite{Carrasco:2015pra,Jang:2015sla,Garron:2016mva,Boyle:2017skn,Boyle:2017ssm}.
For some of these operators, these results turn out to differ again significantly from the earlier results obtained by vacuum insertion approximation (VIA)
\cite{Gerard:1984bg,Gabbiani:1996hi},  with the pattern of deviations summarized 
already in the abstract and discussed in more detail below. To our knowledge 
no attempt has been made so far in the literature to understand the dynamics behind 
this peculiar pattern. The question  thus arises whether DQCD approach could again
help in getting an insight in the lattice QCD results. The main goal of the present paper is the demonstration that this is indeed the case. { This is a
significant result as it underlines the importance of meson evolution in 
a non-leptonic transition in which the controversial role of final state interactions is absent.}

Our paper is organized as follows. In Section~\ref{sec:2} we recall very 
briefly the elements of DQCD relevant for our paper. In Section~\ref{sec:3} 
we give the expressions for the operators $\mathcal{O}_i$, {recall the commonly used parametrization of their $K^0-\bar K^0$ matrix elements} in terms of scale dependent parameters $B_i(\mu)$ and summarize their 
values as obtained by lattice QCD at $\mu=3\gev$  \cite{Carrasco:2015pra,Jang:2015sla,Garron:2016mva,Boyle:2017skn,Boyle:2017ssm}. In  Section~\ref{sec:4} 
we calculate these matrix elements in DQCD. We begin with the large $N$ limit
that corresponds to an appropriate  scale  momentarily denoted by $\mu_0$. 
Next we demonstrate that starting with the lattice QCD results at 
$\mu=3\gev$, performing first renormalization group QCD evolution (quark-gluon evolution) down to scales $\ord(1\gev)$ and subsequently meson evolution in the 
framework of DQCD down to $\mu_0$ {reproduces} rather well the pattern of the 
values of $B_i$ parameters found in the large $N$ limit.
A brief summary is given in  Section~\ref{sec:5}. As we mainly want to 
understand the pattern of the values of $B_i$ in an analytic approach, 
we do not aim for precision 
and perform the $1/N$ meson evolution with the light pseudoscalars only  which is sufficient
for our purposes.

\section{Basics of Dual QCD Approach}\label{sec:2}
The explicit calculation of the contributions of pseudoscalars to hadronic matrix elements of local operators is based on  a truncated chiral Lagrangian describing the low energy 
interactions of the lightest mesons \cite{Chivukula:1986du,Bardeen:1986vp,Bardeen:1986uz}
\be\label{chL}
L_{tr}=\frac{F^2}{8}\left[\text{Tr}(D_\mu UD_\mu U^\dagger)+r\text{Tr}(mU^\dagger+\text{h.c.})-\frac{r}{\Lambda^2_\chi}\text{Tr}(mD^2U^\dagger+\text{h.c.})\right]
\ee
where 
\be\label{UU}
U=\exp(i\sqrt{2}\frac{\Pi}{F}), \qquad 
\Pi=\sum_{\alpha=1}^8\lambda_\alpha\pi^\alpha
\ee
is the unitary chiral matrix describing the octet of light pseudoscalars. 
The parameter $F$ is related to  the weak decay constants $F_\pi\approx 130\mev$ 
and $F_K\approx 156\mev$ through
\be\label{FpiFK}
F_\pi=F\left(1+\frac{m_\pi^2}{\Lambda^2_\chi}\right), \qquad F_K=F\left(1+\frac{m_K^2}{\Lambda^2_\chi}\right),
\ee
so that $\Lambda_\chi\approx 1.1\gev$.
The diagonal mass matrix $m$ involving $m_u$, $m_d$ and $m_s$ is such
that 
\be\label{rr}
r(\mu)=\frac{2 m_K^2}{m_s(\mu)+m_d(\mu)},
\ee
with $r(1\gev)\approx 3.75\gev$ for $(m_s+m_d)(1\gev)\approx 132\mev$.

The flavour-singlet $\eta_0$ meson decouples due to large mass $m_0$ generated 
by the non-perturbative axial anomaly. Consequently the matrix $\Pi$ in (\ref{UU}) reads

\begin{align}\label{eq:Umatrix}
 \Pi = \begin{pmatrix}
                \pi^0+\frac{1}{\sqrt{3}}\eta_8 & \sqrt{2}\pi^+ & \sqrt{2}K^+ \\
		\sqrt{2}\pi^- & -\pi^0+\frac{1}{\sqrt{3}}\eta_8 &  \sqrt{2}K^0 \\
	 \sqrt{2}K^- &  \sqrt{2}\bar K^0 & -\frac{2}{\sqrt{3}}\eta_8
                \end{pmatrix}.
\end{align}
In order to calculate the matrix elements of the local operators in question we 
need meson representations of colour-singlet quark currents and densities. 
They are directly obtained from the effective Lagrangian in (\ref{chL}) and 
are given respectively as follows
\be\label{VAc}
\bar q^b_L\gamma_\mu q^a_L=i\frac{F^2}{8}\left\{(\partial_\mu U)U^\dagger-U(\partial_\mu U^\dagger)+
\frac{r}{\Lambda^2_\chi}\left[(\partial_\mu U)m^\dagger-m(\partial_\mu U^\dagger)\right]\right\}^{ab},
\ee
\be\label{RLd}
\bar q_R^b q_L^a=-\frac{F^2}{8}r\left[U-\frac{1}{\Lambda_\chi^2}\partial^2U\right]^{ab}\,,
\ee
with $U$ turned into $U^\dagger$ under parity.

At the tree level, corresponding to leading order in $1/N$, one uses these representations  to simply express the operators in terms of the meson fields and expands the matrix $U$ in powers of $1/F$. 
For $K^0-\bar K^0$ mixing the relevant contribution to hadronic matrix elements is read off from terms involving only $K^0$ and $\bar K^0$. 
\boldmath
\section{BSM Matrix Elements in Lattice QCD}\label{sec:3}
\unboldmath
There are two equivalent bases for new operators contributing to $K^0-\bar K^0$ mixing. The so-called SUSY basis \cite{Gerard:1984bg,Gabbiani:1996hi} is
 given by
\bea \label{eq:susy}
\mathcal{O}_1 &=& \bar s^\alpha \gamma_\mu P_L d^\alpha \ \bar s^\beta 
\gamma_\mu P_L 
d^\beta\, , \nn \\ 
\mathcal{O}_2 &=& \bar s^\alpha P_L d^\alpha \ \bar s^\beta P_L d^\beta 
\, , \nn \\ 
\mathcal{O}_3&=& \bar s^\alpha P_L d^\beta \ \bar s^\beta P_L  d^\alpha 
\, , \\ 
\mathcal{O}_4 &=& \bar s^\alpha P_L d^\alpha \ \bar s^\beta P_R d^\beta 
\, , \nn \\ 
\mathcal{O}_5&=& \bar s^\alpha P_L d^\beta \ \bar s^\beta P_R d^\alpha
\, , \nn 
\eea 
and the BMU one 
\cite{Buras:2000if}  by
\bea\label{equ:operatorsZ}
{Q}_1^\text{VLL}&=&\left(\bar s\gamma_\mu P_L d\right)\left(\bar s\gamma^\mu P_L d\right) =\mathcal{O}_1 ,\nn \\
{Q}_1^\text{LR}&=&\left(\bar s\gamma_\mu P_L d\right)\left(\bar s\gamma^\mu P_R d\right)=-2\mathcal{O}_5 ,\nn\\
{Q}_2^\text{LR}&=&\left(\bar s P_L d\right)\left(\bar s P_R d\right)=\mathcal{O}_4 \,,\\
{Q}_1^\text{SLL}&=&\left(\bar s P_L d\right)\left(\bar s P_L d\right)=\mathcal{O}_2 \,,\nn\\
{Q}_2^\text{SLL}&=&\left(\bar s \sigma_{\mu\nu} P_L d\right)\left(\bar s\sigma^{\mu\nu}  P_L d\right)=4 \mathcal{O}_2+8 \mathcal{O}_3  \,,\nn
\eea
where 
\be
P_{R,L}=\frac{1}{2}(1\pm\gamma_5), \qquad \sigma_{\mu\nu} = \frac{1}{2} [\gamma_{\mu}, \gamma_{\nu}]\,.
\ee
In (\ref{equ:operatorsZ}) we omitted colour indices as they are summed up in each parenthesis.

In what follows we will use the SUSY basis (\ref{eq:susy}) as most lattice collaborations use it. The 
$K^0-\bar K^0$  matrix elements in this basis are parametrized as follows
\begin{eqnarray}
        \langle \bar K^0| \mathcal{O}_1(\mu)| K^0 \rangle &=&
                        \f{2}{3} m_K^2 F_K^2 B_1 (\mu)\label{MO1} ,\\
         \langle \bar K^0| \mathcal{O}_2(\mu)| K^0 \rangle &=&
                        -\f{5}{12} R(\mu)\, m_K^2 F_K^2 B_2 (\mu) ,\label{MO2}\\
         \langle \bar K^0| \mathcal{O}_3(\mu)| K^0 \rangle &=&
                        \f{1}{12} R(\mu)\, m_K^2 F_K^2 B_3 (\mu) ,\label{MO3}\\
         \langle \bar K^0| \mathcal{O}_4(\mu)| K^0 \rangle &=&
                        \f{1}{2} R(\mu)\, m_K^2 F_K^2 B_4(\mu) ,\label{MO4}\\
         \langle \bar K^0| \mathcal{O}_5(\mu)| K^0 \rangle &=&
                        \f{1}{6} R(\mu)\, m_K^2 F_K^2 B_5 (\mu),\label{MO5}
\end{eqnarray}
where 
\be
        R(\mu) = \left( \f{m_K}{m_s(\mu) + m_d(\mu)} \right)^2 = \frac{r^2(\mu)}{4 m_K^2}
\ee
refers to the generic factorized evolution of any density-density operator with $\mu$, a renormalization scale taken at 
$3\gev$ by  ETM, SWME and RBC-UKQCD lattice  collaborations.

Still the BMU basis in (\ref{equ:operatorsZ}) will turn out to be very useful at 
intermediate steps because all its  operators are made of colour singlet bilinears {that are most suitable for DQCD 
calculations.

Now in the vacuum insertion approximation (VIA) the parameters $B_i$ are,  by definition,  given by
\be\label{VIA}
B_1=B_2=B_3=B_4=B_5=1 \,\qquad {(\rm VIA)}
\ee
and thus $\mu$-independent.
Already this property of VIA, which is based on the factorization of matrix elements
of four-quark operators into products of quark currents or quark densities, 
is problematic. The only way out would be that VIA is valid only at a single 
scale. The problem is that VIA by itself does not specify this scale and, as
we will see soon, there is no value of $\mu$ at which the relations in (\ref{VIA}) are satisfied in QCD.

In the DQCD approach on the other hand, the factorization of matrix elements 
in question can be proven to be a property of QCD in the large $N$ limit
\cite{Bardeen:1986vp,Bardeen:1986uz,Bardeen:1986vz,Bardeen:1987vg} because 
in this limit  QCD at very low momenta becomes a free theory 
of mesons \cite{'tHooft:1973jz,'tHooft:1974hx,Witten:1979kh,Treiman:1986ep}.
 With  non-interacting mesons the factorization of matrix elements 
of four-quark operators into matrix elements of quark currents and quark 
densities is automatic. But even then several $B_i$ parameters are not equal to 
unity as VIA includes the so-called Fierz-terms that are $1/N$ suppressed 
and thus absent in the large $N$ limit. The classic example is the parameter $B_1$ which is equal to unity in the VIA and to $3/4$  in the large $N$ limit 
 \cite{Buras:1985yx}.

But another advantage of DQCD over VIA is that it tells us that factorization 
in question 
does not take place at values of $\mu$ used by lattice QCD collaborations but rather at zero momentum transfer between colour-singlet currents or densities. 
 Therefore it is not surprising that lattice QCD 
collaborations find  $B_1(\mu)\approx 0.53$ at $\mu=3\gev$, a value significantly 
below the large $N$ one, i.e., $3/4$. While within numerical approach like lattice QCD this 
difference cannot be understood, this decrease of $B_1(\mu)$ with increasing 
$\mu$ can be shown
to be the consequence of meson evolution from scale $\mu_0=0$  to 
scale $\ord(1\gev)$ followed by the usual renormalization group {quark-gluon} evolution
up to $\mu=3\gev$  \cite{Bardeen:1987vg}. In this particular case,
one usually multiplies the result
 by the corresponding SD  renormalization group factor to find the scale and renormalization scheme independent  $\hat B_K=0.73\pm0.02$  \cite{Buras:2014maa} in a very good 
agreement with the  world average of lattice QCD calculations  
$\hat B_K=0.766\pm 0.010$ \cite{Aoki:2016frl}. 

In the case of the BSM operators $\mathcal{O}_i$ with $i=2-5$ the construction 
of scale independent $\hat B_i$ parameters, although possible, is not particular
useful because  $\mathcal{O}_2$ mixes under renormalization with $\mathcal{O}_3$
and $\mathcal{O}_4$ with  $\mathcal{O}_5$. This mixing is known at the NLO level
 \cite{Ciuchini:1997bw,Buras:2000if} and useful NLO expressions for $\mu$ dependence of hadronic matrix elements and their Wilson coefficients can be found in 
\cite{Buras:2001ra}. But for our 
purposes it will be sufficient to work in LO approximation and use LO formulae 
also given in \cite{Buras:2001ra}.

The lattice QCD results from three collaborations are displayed in 
Table~\ref{LATTICE}.
\begin{table}[t]
\centering
\renewcommand{\arraystretch}{1.4}
\begin{tabular}{|c|c|c|c|c|c|}
\hline
 Collaboration
& $n_f$ &$B_2$ & $B_3$ & $B_4$ & $B_5$
\\
\hline
ETM15 & $2+1+1$ & $0.46(3)(1)$ & $0.79(5)(1)$ & $0.78(4)(3)$ & $0.49(4)(1)$
\\
SWME15 & $2+1$ &  $0.525(1)(23)$ & $0.773(6)(35)$ & $0.981(3)(62)$ & $0.751(7)(68)$
\\
RBC-UKQCD & $2+1$ & $0.488(7)(17)$ & $0.743(14)(65)$ & $0.920(12)(16)$ & $0.707(8)(44)$
\\
\hline
\end{tabular}
\renewcommand{\arraystretch}{1.0}
\caption{\label{LATTICE}
  Results for the parameters $B_i$ with the first error being statistical and the second systematic obtained by ETM \cite{Carrasco:2015pra},
 SWME \cite{Jang:2015sla} and RBC-UKQCD \cite{Garron:2016mva,Boyle:2017skn,Boyle:2017ssm} collaborations at $\mu=3\gev$. Useful comparison can be found in
\cite{Boyle:2017ssm}.
}
\end{table}

We observe that:
\begin{itemize}
\item
there is a good agreement between three collaborations as far as the values of $B_2$ and $B_3$ are concerned;
\item
the value of $B_4$ obtained by SWME15 and RBC-UKQCD are close to unity, while 
the ETM collaboration obtains a significantly lower value;
\item
the value of $B_5$ obtained by SWME15 and RBC-UKQCD are in the ballpark of 
$0.7$, while the value from ETM collaboration is in the ballpark of $0.5$;
\item
 most values differ significantly from unity, prohibiting the use 
of VIA. 
\end{itemize}

To our knowledge no lattice group made an attempt to understand this peculiar
pattern of values. On the other hand, as we will demonstrate now, it can be
understood within DQCD  because in this approach an insight in 
the QCD dynamics at very low scales up to $ 1\gev$ can be obtained through 
{\it the  meson evolution} followed by the usual RG QCD evolution, termed in 
\cite{Bardeen:1986vp,Bardeen:1986uz,Bardeen:1986vz,Bardeen:1987vg,Buras:2014maa}
 {\it the quark-gluon evolution,} from  $1\gev$ to $3\gev$.

\boldmath
\section{Calculating BSM $B_i$ in DQCD}\label{sec:4}
\subsection{Large $N$ Limit}
\unboldmath
Using the meson representation of quark densities in (\ref{RLd}), the $K^0-\bar K^0$
 matrix elements of the local operators  $\mathcal{O}_2$ and $\mathcal{O}_4$ 
can be easily calculated in the 
 large $N$ limit because these operators are built out of colour singlet quark densities. We find
\be
 \langle \bar K^0| \mathcal{O}_4(\mu_0)|K^0 \rangle_\infty\, = -\langle\bar K^0| \mathcal{O}_2(\mu_0)|K^0 \rangle_\infty =
 \f{1}{2} R(\mu_0)\, m^2_K F_K^2 
\ee
where we used the relations (\ref{FpiFK}). Comparing with (\ref{MO2}) and (\ref{MO4}) and interpreting the factorization to be valid at $\mu_0=0$, we extract
\be\label{B24}
B_2(0)=1.20, \qquad  B_4(0)=1.00 \qquad ({\rm large~N~limit})\,.
\ee
We note that $B_2\not=1$ because a $1/N$ Fierz term included in VIA is absent
now.

In order to find the remaining two matrix elements we have to bring them to the 
colour singlet form with the help of {relations in (\ref{equ:operatorsZ}).} In the case of  $\mathcal{O}_3$  we use  
the fact that  $\langle  {Q}_2^\text{SLL}(\mu_0)\rangle=0$ 
to obtain 
\be\label{B33}
\langle \bar K^0| \mathcal{O}_3(\mu_0)|K^0 \rangle_\infty = -\frac{1}{2} 
\langle \bar K^0| \mathcal{O}_2 (\mu_0)|K^0\rangle_\infty = \f{1}{4} R(\mu_0)\, m^2_K F_K^2 \,,
\ee
namely, from (\ref{MO3}),
\be\label{B3}
 B_3(0)=3.0   \qquad  \qquad ({\rm large~N~limit}).
\ee

Similarly, for the matrix element of $ \mathcal{O}_5$ expressed in terms 
of the currents in (\ref{VAc}) 
we find using  {(\ref{equ:operatorsZ})}
\be
 \langle \bar K^0| \mathcal{O}_5(\mu_0)|K^0 \rangle_\infty = -\f{1}{2}
 \langle \bar K^0| {Q}_1^\text{LR}(\mu_0)|K^0 \rangle_\infty =
\frac{1}{4}\, m^2_K F_K^2\,
 \ee
and extract,  using (\ref{MO5}),
\be\label{B50}
R(\mu_0)  B_5(\mu_0)=\frac{3}{2}\,.
\ee

The result for $B_5$ is rather peculiar, which is related to the fact that in the BMU basis $\mathcal{O}_5$ is not a density-density operator but a
current-current  one like $\mathcal{O}_1$. This implies that the usual parametrization of 
 $\langle\mathcal{O}_5\rangle$ as given in (\ref{MO5}) in terms of $B_5$ 
is not useful at 
low scales where $R(\mu)$ is not accessible. But in order to compare it later with lattice QCD 
we estimate its value by 
evaluating $R\approx 6.5$ at the scale around $0.7\gev$ where the meson evolution matches the quark-gluon one for the $\Delta S$ = 1 operators \cite{Buras:2018evv}, to find
\be\label{B5}
B_5(0)\approx 0.23    \qquad ({\rm ~large~N~limit})\,.
\ee
The main message from this estimate is that  at low scales $B_5$ is expected to be smaller than 
the remaining $B_i$ parameters.

In any case we observe that, except for $B_4$, the results for parameters 
$B_i$ obtained in the 
large $N$ limit differ significantly from the Lattice ones in Table~\ref{LATTICE} as well as from VIA in (\ref{VIA}). 

As the lattice results and large $N$ results correspond to quite different scales,
in order to understand lattice results with 
the help of DQCD  we will proceed in two steps as follows:
\begin{itemize}
\item
 we will start with the lattice QCD values for $B_i$ at $\mu=3\gev$  in Table~\ref{LATTICE} and evolve these parameters through quark-gluon evolution down to the scale $\mu=1\gev$;
\item 
in order to see whether the values $B_i(1\gev)$ from LQCD can be explained in 
DQCD, we will start with large $N$ values in (\ref{B24}), (\ref{B3}) and 
(\ref{B5}) and perform the meson evolution up to  
$\Lambda=(0.65\pm0.05)\gev$. 
\end{itemize}

Ideally one would like to perform meson evolution up to  $\mu=1\gev$
but, as only pseudoscalar contributions are taken into account, one has
to stop the evolution around $0.7\gev$ so that the comparison will not
be perfect but sufficient to reach firm conclusions by extrapolation.

In this context we should recall that 
the parameters $\bsi$ and $\bei$ relevant for $K\to\pi\pi$ decays 
equal unity in the large $N$ limit to be compared with the 
values  in the ballpark of $0.6$ and $0.8$, respectively, obtained by RBC-UKQCD collaboration at $\mu=1.5\gev$. This pattern 
can be understood at least semi-quantitatively in DQCD  
\cite{Buras:2015xba,Buras:2016fys} and the question arises, 
whether the pattern 
\be\label{pattern1}
B_2 < B_5\le B_3 < B_4 \, \qquad (\, {\rm at}~\mu= 3\gev)
\ee
can be understood within this framework as well, although in DQCD one
finds 
\be\label{pattern2}
B_5 \ll  B_4 < B_2 < B_3\, \qquad (\,{\rm at}~\mu\ll 1\gev)\,.
\ee

\subsection{Quark-Gluon Evolution}
Let us first check whether already 
the perturbative evolution from  {\bf $\mu=3\gev$ to $\mu=1\gev$ } allows us
to understand partially the difference between the patterns (\ref{pattern1})
and (\ref{pattern2}). In order to disentangle the non-factorizable $B_i(\mu)$ from the factorizable $R(\mu)$, we work
 in the one-loop approximation.
The
anomalous dimension matrices are then given in units of $\as/4\pi$ as follows:
\bea \label{g023}
\hat{\gamma}^{(0)}(\mathcal{O}_2,\mathcal{O}_3) &=& \left( \begin{array}{ccc} 
-{6}{N}+8+\f{2}{N} &~& 4-\f{8}{N} \\[1mm]
4 N-4-\f{8}{N} && 2 N+4+\f{2}{N} \end{array} \right),
\\[2mm]
\hat{\gamma}^{(0)}(\mathcal{O}_4,\mathcal{O}_5) &=& \left( \begin{array}{ccc}
-6 N +\f{6}{N} &~& 0 \\[1mm]
-6    && \f{6}{N} 
\end{array} \right). \label{g045} 
\eea
\noindent
The two operator systems $(\mathcal{O}_2,\mathcal{O}_3)$ and $(\mathcal{O}_4,\mathcal{O}_5)$ do not mix under renormalization with each other. 

Now keeping only first leading logarithms one has for $\mu_2>\mu_1$
\be\label{RGOi}
\langle \mathcal{O}_i(\mu_2)\rangle= 
\langle\mathcal{O}_i(\mu_1)\rangle\left(1-\aspi {\hat{\gamma}^{(0)}_{ii}}\ln(\f{\mu_2}{\mu_1})\right) -
\langle \mathcal{O}_j(\mu_1)\rangle\aspi {\hat{\gamma}^{(0)}_{ij}}\ln(\f{\mu_2}{\mu_1})\,.
\ee

Taking into account that the factorized $R(\mu)$ scales like
\be\label{RMU}
\f{R(\mu_2)}{R(\mu_1)}=\left[1+2\aspi\gamma_m^{(0)}\ln(\f{\mu_2}{\mu_1})\right]
,\qquad  \gamma_m^{(0)}=6\,\left(\f{N^2-1}{2N}\right)
\ee
and inserting the expressions for $\langle \mathcal{O}_i(\mu)\rangle$ in
(\ref{MO2}-\ref{MO5}) into (\ref{RGOi}), we isolate in this manner the $\mu$ dependence of 
$B_i(\mu)$ 
\be\label{LO1}
B_2(\mu_2)=B_2(\mu_1)\left[1-\f{(8N-4)}{N}\aspi\ln(\f{\mu_2}{\mu_1})\right]+
B_3(\mu_1)\f{(4N-8)}{5N}\aspi\ln(\f{\mu_2}{\mu_1})\,,
\ee
\be\label{LO2}
B_3(\mu_2)=B_3(\mu_1)\left[1-P_{33}(N)\aspi\ln(\f{\mu_2}{\mu_1})\right]+
B_2(\mu_1) P_{32}(N)\aspi\ln(\f{\mu_2}{\mu_1})\,.
\ee
with 
\be
P_{33}(N)=\f{(8N^2+4N-4)}{N}\,, \qquad P_{32}(N)=\f{(20 N^2-20N-40)}{N}
\ee
and
\be\label{LO3}
B_4(\mu_2)=B_4(\mu_1)\,,~~~~~~~~~~~~~~~~~~~~~~~~\qquad\qquad~~~~~~~~~~~~~~~~~~~
\ee
\be\label{LO4}
B_5(\mu_2)=B_5(\mu_1)\left[1-6N\aspi\ln(\f{\mu_2}{\mu_1})\right]+
B_4(\mu_1)18\aspi\ln(\f{\mu_2}{\mu_1})\,.
\ee

We observe: 

\begin{itemize}
\item
 a strong suppression of $B_2$ with increasing $\mu$ due to the first term 
in (\ref{LO1});
\item
an even stronger  suppression of $B_3$ with $B_3>B_2$ due to the first term in (\ref{LO2});
\item
no $\mu$ dependence of $B_4$ in (\ref{LO3});
\item
a positive shift of $B_5$ with $B_5\ll B_4$, due to the second term in (\ref{LO4}).
\end{itemize}

However, in view of large coefficients in front of the logarithms for $N=3$ one should  sum them to all orders.  We do this in Appendix~\ref{SLL}.
This allows  us to find the values of $B_i$ for $\mu_1=1 \gev$ from 
those obtained by RBC-UKQCD at $\mu_2=3\gev$ and given in Table~\ref{LATTICE}.
 Inserting the central lattice values in the formulae of Appendix~\ref{SLL}, we 
find at $\mu=1\gev$
\be\label{L1GEV}
B_2=0.608,\qquad B_3=1.06, \qquad B_4=0.920, \qquad B_5=0.519 \,.
\ee
The corresponding values obtained from central values of ETM and SWME 
collaborations are collected in Table~\ref{tab:epe-LQ-contr}.

As already observed from (\ref{LO1}- \ref{LO4}),
$B_2$, $B_3$ and $B_5$, all moved towards their large $N$ 
values in (\ref{B24}), (\ref{B3}) and (\ref{B5}) while $B_4$ did not change in LO approximation. 
These results are already very encouraging.
We will next estimate whether the meson evolution can do the rest of 
the job within DQCD.

\subsection{Meson Evolution}
We first calculate the $1/N$ meson evolution in question in the chiral limit using 
the technology developed in \cite{Fatelo:1994qh}, namely expanding 
\be\label{FG}
\tilde U\equiv \exp(i\sqrt{2}\frac{\Xi}{F})\, U, \qquad \Xi=\sum_{\alpha=1}^8\lambda_\alpha \xi^\alpha
\ee
around the classical field $U$.
   
 It should be emphasized that we do not aim here at achieving
the precision of lattice QCD. We would like mainly to demonstrate that DQCD allows us to understand lattice
results at the semi-quantitative level by including analytically QCD dynamics
at very low scales. It is this dynamics which is responsible for the pattern
in Table~\ref{LATTICE}.

For the non-factorizable evolution of 
density-density operators like $\mathcal{O}_{2,3,4}$,  we find
\be\label{ME3}
(U)^{ab}(U)^{cd}(\Lambda)=(U)^{ab}(U)^{cd}(0)- 4\,\frac{\Lambda^2}{(4\pi F)^2} \left[(U)^{ad}(U)^{cb}-\frac{1}{3}(U)^{ab}(U)^{cd}\right](0)
\ee
and
\be\label{ME4}
(U)^{ab}(U^+)^{cd}(\Lambda)=(U)^{ab}(U^\dagger)^{cd}(0)+ 4\frac{\Lambda^2}{(4\pi F)^2}
\left[\delta^{ad}(U^\dagger U)^{cb}-\frac{1}{3}(U)^{ab}(U^\dagger)^{cd}\right](0)\,.
\ee
The last terms proportional to $1/3$ arise from the Fierz relation  
\be
(\lambda_\alpha)^{ij} (\lambda^\alpha)^{kl} =2 \left(\delta^{il}\delta^{kj}-\frac{1}{3}
\delta^{ij}\delta^{kl}\right)
\ee
for the Gell-Mann matrices introduced in (\ref{UU}) and (\ref{FG}). Had we worked in the nonet approximation, these terms would be absent. In particular, the $\Lambda^2$ term
in (\ref{ME4}) would vanish in remarkable accordance with the absence of
short distance evolution of $B_4$ in the leading log approximation { as 
seen in (\ref{LO3})}.
Consequently, in our DQCD approach the purely non-perturbative contribution from the strong anomaly 
is responsible for $B_4  < 1$. It would be interesting if lattice QCD simulations confirm this feature one day.

The evolution in (\ref{ME3}) can be directly applied to $B_2$. However, the case of $B_3$ is more subtle. Working in the BMU basis (\ref{equ:operatorsZ})
to extract its value in the large $N$ limit, we have to stay in the same basis to study the meson evolution of $\mathcal{O}_3$  since otherwise the running of $R(\mu)$ and $B_i(\mu)$ would be interchanged under Fierzing. Consequently, the meson evolution of  $\mathcal{O}_3$  is governed by a well-defined linear combination of 
${Q}_1^\text{SLL}$ and ${Q}_2^\text{SLL}$, namely
\be
\mathcal{O}_3=-\f{1}{2}\left[{Q}_1^\text{SLL}-\f{1}{4}{Q}_2^\text{SLL}\right]\,.
\ee

On the one hand, the ${Q}_1^\text{SLL}$ evolution down to  the factorization scale $\mu_0$  is fully given by (\ref{ME3}) since $\langle{Q}_2^\text{SLL}(\mu_0)\rangle=0$.  On the other, to infer the non-trivial meson evolution of ${Q}_2^\text{SLL}$  into ${Q}_1^\text{SLL}$  above $\mu_0$ we have to rely on the mixing pattern of the quark-gluon anomalous dimension matrix {given in units of $\alpha_s/4\pi$ in the BMU basis at one-loop  as follows
\bea \label{g012}
\hat{\gamma}^{(0)}({Q}_1^\text{SLL},{Q}_2^\text{SLL}) &=& \left( \begin{array}{ccc} 
-{6}{N}+\f{6}{N}+6 &~& \f{1}{2}-\f{1}{N} \\[1mm]
 -24-\f{48}{N} && 2 N-\f{2}{N} +6 \end{array} \right),
\eea   

For that purpose, we omit again the $(- 6N + 6/N)$ term in the $\hat{\gamma}^{(0)}_{11}$ entry since it corresponds to the running of the factorized $R(\mu)$
given in (\ref{RMU}). Extending then this mixing pattern below $1\gev$, we expect the left-over relative factor $(- 4)$ between $\hat{\gamma}^{(0)}_{21}$ and $\hat{\gamma}^{(0)}_{11}$ to survive hadronization in the large $N$ and chiral limits. If such is the case, following DQCD, the non-factorizable meson evolution of $\mathcal{O}_3(\Lambda^2)$ down into  
$\mathcal{O}_2(0)$ is in fact a factor of two faster than the one for 
$\mathcal{O}_2$ so that the factor $-4$ in (\ref{ME3}) is replaced by $-8$.

For the evolution of local current-current operators like  $\mathcal{O}_5$ in the BMU basis, we obtain
\be
(J_L)^{ab}(J_R)^{cd}(\Lambda)=(J_L)^{ab}(J_R)^{cd}(0)+ \frac{\Lambda^4}{(4\pi F)^2}\left[\frac{F^4}{8}\right]  \left[(U)^{ad}(U^\dagger)^{cb}-\frac{1}{3}\delta^{ab}(U^\dagger U)^{cd}\right](0).
\ee

Requiring the measure to be chiral invariant, such a quartic cut-off dependence can be properly cancelled \cite{Fatelo:1994qh}. Computing the quadratic one in the specific case of a Fierz-conjugate $\Delta S = 2$ transition we find then:
\be\label{JLJR}
(J_L)^{ds}(J_R)^{ds}(\Lambda)=(J_L)^{ds}(J_R)^{ds}(0)-\frac{\Lambda^2}{(4\pi F)^2}\left[\frac{F^4m_K^2}{4}\right](U)^{ds}(U^\dagger)^{ds}(0)\,.
\ee

Starting with the large $N$ values in  (\ref{B24}), (\ref{B3}) and (\ref{B5})
 and letting them evolve on the basis of (\ref{ME3}), (\ref{ME4}) and (\ref{JLJR})
we find at order $1/N$
\be\label{R1}
B_2(\Lambda)=1.2\, \left[1 - \frac{8}{3}\, \f{\Lambda^2}{(4\pi F_K)^2}\right],
\ee
\be\label{R2}
B_3(\Lambda)=3.0\, \left[1 - \frac{16}{3}\,  \f{\Lambda^2}{(4\pi F_K)^2} \right],
\ee
\be\label{R3}
B_4(\Lambda)=1.0\, \left[1 -  \frac{4}{3}\,  \f{\Lambda^2}{(4\pi F_K)^2}\right]
\ee
\be\label{R4}
B_5(\Lambda) = 0.23\,
\left[1 + 4\,  \f{\Lambda^2}{(4\pi F_K)^2}\right]\,\,,
\ee
where $\Lambda$ is the cut-off of DQCD which allows us to separate the non-factorizable meson evolution from the quark-gluon one.

The general trend already observed in the quark-gluon evolution is nicely outlined with
\begin{itemize}
\item
a strong suppression of $B_2$;
\item
an even stronger suppression of $B_3$;
\item
a smooth evolution of $B_4$;
\item
a strong enhancement of $B_5$.
\end{itemize}

In the chiral limit  we work and including only pseudoscalar contributions in the loops, a reasonable range for  $\Lambda$ is 
\be
m_8\le \Lambda < m_0
\ee
with
\be
m_8^2=\frac{4m_K^2-m^2_\pi}{3}\approx (0.57\gev)^2
\ee
the Gell-Mann-Okubo mass relation in the octet approximation, and
\be
m_0^2=m_\eta^2+m^2_{\eta^\prime}-2 m_K^2 \approx (0.85\gev)^2
\ee
the axial anomaly mass relation 
in the large $N$ limit \cite{Gerard:2004gx}. In the nonet approximation
(i.e., $m_0=0$), the factors $(-8/3,-16/3,-4/3,+4)$ in  (\ref{R1}-\ref{R4}) would be replaced by $(-4,-8,0,+4)$, respectively.

Including vector meson contributions  would allow us to raise the cut-off $\Lambda$ and to approach the critical scale of $1\gev$. 
But we do not think such a complication is necessary in order 
to explain the pattern found by lattice QCD.

Indeed, setting  $\Lambda=0.6\,(0.7)\gev$   in  (\ref{R1}-\ref{R4}),
we obtain
\be
B_2(\Lambda)\approx 0.9\,(0.8)\, \quad  B_3(\Lambda)\approx 1.5\,(1.0)\,  \quad  B_4(\Lambda)\approx 0.9\,(0.8)\,\quad B_5(\Lambda)\approx 0.29\,(0.35)\,.
\ee

 This should be compared with lattice QCD values in (\ref{L1GEV}). Again correct pattern is reproduced and it is justified to conclude that these results around $0.65\gev$ are already satisfactory.

The $\mathcal{O}_1$ and $\mathcal{O}_5$ current-current operators do not receive one-loop corrections from the flavour-singlet $\eta_0$ meson. So, we can safely
consider the full chiral corrections given in Appendix~\ref{BCHIRAL} for them
to  obtain the final results for $\Lambda=(0.65\pm0.05)\gev$ given in Table~\ref{tab:epe-LQ-contr}.

Let us recall that  in the case of the parameter $\hat B_K$ or equvalently 
$B_1$ and in the case of $K\to\pi\pi$ decays within the SM we have  
included not only pseudoscalar meson contributions, as done here, but also those
of lightest vector mesons. As seen in Tables 4 and 5 in \cite{Buras:2014maa} this improvement has an impact on final numerical values and 
improves  the matching of both evolutions but, importantly for us,
does not modify the signs of evolutions. We expect 
this to be also the case here.

Our results, collected in Table~\ref{tab:epe-LQ-contr}, demonstrate very clearly that the pattern 
of $B_i$  values obtained by QCD lattice collaborations  can be understood in 
DQCD by performing quark-gluon evolution of lattice values down to $\mu=1\gev$ 
and the meson evolution from the factorization scale up to scales $\ord(1\gev)$.

\begin{table}[t]
\centering
\renewcommand{\arraystretch}{1.4}
\begin{tabular}{||c|c|c|c|c||}
\hline
 $\mu$
& $B_2$ & $B_3$ & $B_4$ & $B_5$
\\
\hline
$3\gev$ (ETM15) &$0.46$ & $0.79$ & $0.78$ & $0.49$\\
$3\gev$ (SWME) &$0.52$ & $0.77$ & $0.98$ & $0.75$\\
$3\gev$ (RBC-UKQCD) &$0.49$ & $0.74$ & $0.92$ & $0.71$
\\
$\Downarrow$ (QG) & & & & \\
$1\gev$ (ETM15) & $0.58$& $1.22$ & $0.78$ & $0.24$ \\
$1\gev$ (SWME) & $0.66$& $1.07$ & $0.98$ & $0.55$ \\
$1\gev$ (RBC-UKQCD) & $0.61$& $1.06$ & $0.92$ & $0.52$ \\
\hline\hline
$(0.60\to 0.70)\gev$ &  $0.90\to 0.79$& $1.50\to 0.96$ & $0.87\to 0.83$ & $0.27\to 0.30$\\
$\Uparrow$ (DQCD) & & & & \\
$0$ & 1.2 & 3.0 & 1.0 & 0.23\\
\hline 
 $\Lambda$
& $B_2$ & $B_3$ & $B_4$ & $B_5$ \\
\hline
\end{tabular}
\renewcommand{\arraystretch}{1.0}
\caption{\label{tab:epe-LQ-contr}
  When mesons (almost) meet quarks and gluons. Upper half: summary of the central values of $B_i$ parameters at $\mu=3\gev$ obtained by  Lattice QCD collaborations and their quark-gluon  (QG) evolution down to $\mu=1\gev$. Lower half:  summary of the values of $B_i$ parameters obtained in DQCD  and their meson evolution 
up to $\Lambda=(0.60\to 0.70)\gev$. 
}
\end{table}

{
In judging the quality of the agreement of DQCD with Lattice QCD we should 
realize that there is still a gap between the values of the cut-off $\Lambda$ 
and $\mu=1\gev$ at which the values of $B_i$ obtained by lattice QCD can be
evaluated. Moreover, except for $B_5$ our calculations have been performed 
in the chiral limit. Going beyond this limit and also including the vector
meson contributions would allow us to improve the calculation and raise 
the values of $\Lambda$ at least to $0.9\gev$. 

Extrapolating Table~\ref{tab:epe-LQ-contr} with some caution, we can make the
following observations.
\begin{itemize}
\item
The parameter $B_2$ is still visibly larger than lattice values at $1\gev$ 
but the difference between the lattice values at $3\gev$ and its large $N$ limit
value 
of roughly 0.7 has been decreased below 0.2 for $\Lambda =0.70\gev$. It 
would disappear already for  $\Lambda =0.80\gev$.
\item
The parameter $B_3$ is already in a very good agreement with lattice values, 
although its value is subject to visible uncertainty. Yet, what is impressive is
that the initial difference between DQCD and lattice QCD values of $2.2$ has 
been decreased to $0.2$ at  $\Lambda =0.70\gev$.
\item
The meson evolution decreases $B_4$ by roughly $15\%$.
 At $\Lambda=0.70\gev$ we find the value of $B_4$ between those obtained by 
ETM collaboration and RBC-UKQCD.  It will be interesting to see how this
comparison will look like when the accuracy of lattice results will improve.
\item
Our result for $B_5$ is between those obtained by 
ETM and the two other collaborations.
\end{itemize}

}

\section{Summary}\label{sec:5}

While  not as precise as ultimate lattice QCD calculations, the DQCD approach offered over  many years an insight in the lattice results and often, like was the case
of the $\Delta I=1/2$ rule \cite{Bardeen:1986vz} and the parameter $\hat B_K$ \cite{Bardeen:1987vg}, provided results 
 almost three decades before this was possible with lattice QCD. 
The agreement between results from DQCD approach and lattice QCD is
remarkable, in particular considering the simplicity of the former approach 
and the very sophisticated and tedious numerical calculations of the latter. The most recent example is the good agreement for hadronic 
matrix elements of the chromomagnetic penguin operator between DQCD \cite{Buras:2018evv} and results from the ETM collaboration \cite{Constantinou:2017sgv}.}

In the present paper we have demonstrated that the pattern of 
hadronic matrix elements of BSM operators affecting $K^0-\bar K^0$ mixing 
obtained by lattice QCD can also be understood in DQCD at the semi-quantitative level. {The crucial role in this insight, as seen in Table~\ref{tab:epe-LQ-contr}, is played by meson evolution, an important ingredient of DQCD, which could be exhibited here clearer 
than in $K\to\pi\pi$ decays because of the absence of final state interactions
 in $K^0-\bar K^0$ mixing. In turn our 
results imply that the inclusion of 
 meson evolution in the phenomenology of $K\to\pi\pi$ decays is mandatory.}

It is truly remarkable that this insight has been obtained without
any free parameters beyond the value of the cut-off $\Lambda$. The remaining 
input were the values of the pseudoscalar masses, $F_K$ and of $\alpha_s$. 
In particular no values of low-energy constants from lattice QCD, used
often in chiral perturbation studies were involved.
We are not aware of any analytical approach that could provide such 
insight in lattice QCD results in question. This makes us confident that the 
pattern for  $\bsi$ and $\bei$ obtained in DQCD \cite{Buras:2015xba,Buras:2016fys} will be confirmed by improved lattice calculations making the existing 
$\epe$ anomaly more pronounced.

\section*{Acknowledgements}
We thank Christoph Bobeth, Peter Boyle, Nicolas Garron and Julia Rachael Kettle for discussions.
This research was supported by the DFG cluster
of excellence ``Origin and Structure of the Universe''.

\appendix

\section{Summing Leading Logarithms}\label{SLL}
In order to sum the leading logarithms 
the formulae (4.4)-(4.7) and (4.12)-(4.15) of
\cite{Buras:2001ra} derived in the BMU basis 
 (\ref{equ:operatorsZ}) have to be transferred into the SUSY one (\ref{eq:susy}).

 We find now
\be
B_2(\mu_2)=T(\mu_2,\mu_1) \left[B_2(\mu_1)r_{22}-\f{1}{5} B_3(\mu_1)r_{23}\right]\,,
\ee
\be
B_3(\mu_2)=T(\mu_2,\mu_1)\left[ B_3(\mu_1)r_{33}-5 B_2(\mu_1) r_{32}\right]\,,
\ee
and
\be
B_4(\mu_2)=T(\mu_2,\mu_1)\left[B_4(\mu_1) r_{44}+\f{1}{3} B_5(\mu_1) r_{45}\right]\,,
\ee
\be
B_5(\mu_2)= T(\mu_2,\mu_1)\left[B_5(\mu_1) r_{55}+3 B_4(\mu_1)r_{54}\right]\,.
\ee
Here
\be\label{TR}
T(\mu_2,\mu_1)=\f{R(\mu_1)}{R(\mu_2)}\left[\f{\alpha_s^{(4)}(\mu_2)}{\alpha_s^{(4)}(m_c)}\right]^{24/25}
\,\left[\f{\alpha_s^{(3)}(m_c)}{\alpha_s^{(3)}(\mu_1)}\right]^{8/9}
\ee
and the coefficients $r_{ij}$ are just linear combinations of the coefficients 
 $\rho_{ij}$ for which explicit expressions can be found 
in (4.4)-(4.7) and (4.12)-(4.15) of \cite{Buras:2001ra}. The $\rho_{ij}$ are just
functions of QCD factors like the ones in (\ref{TR}).

We find
\be
r_{22}=\left[\rho_{11}\right]_{\rm SLL}+4\left[\rho_{21}\right]_{\rm SLL}\,, \qquad
r_{23}= 8 \left[\rho_{21}\right]_{\rm SLL}\,, \qquad
r_{33}=\left[\rho_{22}\right]_{\rm SLL}-4\left[\rho_{21}\right]_{\rm SLL}\,, 
\ee
\be
r_{32}= \frac{1}{2} (\left[\rho_{22}\right]_{\rm SLL}-\left[\rho_{11}\right]_{\rm SLL})-2 \left[\rho_{21}\right]_{\rm SLL} +\frac{1}{8}  \left[\rho_{12}\right]_{\rm SLL}  \,,
\ee
and
\be
r_{44}=\left[\rho_{22}\right]_{\rm LR}\, \qquad r_{45}=-2\left[\rho_{12}\right]_{\rm LR}\,,
\ee
\be
r_{55}=\left[\rho_{11}\right]_{\rm LR}\, \qquad r_{54}=-\frac{1}{2}\left[\rho_{21}\right]_{\rm LR}\,.
\ee
We have suppressed the scales $\mu_K=\mu_1$ and $\mu_L=\mu_2$ as 
well the superscript $(0)$ which indicates LO approximation. In fact the
formulae above are valid including NLO corrections given in that paper except
that the formula  (\ref{TR})  must be then generalized beyond LO. For our 
purposes LO is sufficient.

Using then
\be
\alpha_s^{(4)}(3\gev)=0.241, \qquad 
\alpha_s^{(4)}(m_c)= \alpha_s^{(3)}(m_c)=0.365,\qquad  \alpha_s^{(3)}(1\gev)=0.437
\ee
we find
\be
B_2(\mu_2)=0.794\, B_2(\mu_1)+ 0.005B_3(\mu_1)\,,
\ee
\be
B_3(\mu_2)=0.395\, B_3(\mu_1)+0.532\, B_2(\mu_1)\,,
\ee
and
\be
B_4(\mu_2)=B_4(\mu_1)\,,~~~~~~~~~~\qquad\qquad~~~~~
\ee
\be
B_5(\mu_2)=0.532\, B_5(\mu_1)+ 0.468\, B_4(\mu_1)\,.
\ee
These results generalise the ones in (\ref{LO1})-(\ref{LO4}) to include 
the summation of leading logarithms to all orders in perturbation theory.
Using the relation for three quark theory
\be
\f{\alpha_s(\mu_2)}{\alpha_s(\mu_1)}=1-9\f{\alpha_s}{2\pi}\ln\f{\mu_2}{\mu_1}
\ee
and keeping only the first leading logarithm one verifies the results in 
(\ref{LO1})-(\ref{LO4}).

\section{Meson evolution beyond chiral limit}\label{BCHIRAL}
For the $\mathcal{O}_1$ and $\mathcal{O}_5$ current-current operators, 
chiral corrections imply
\be
B_i(\Lambda)=B_i(0)\left(1+\f{\Delta_i(\Lambda)}{2F_K^2}\right)
\ee
with
\be
\begin{aligned}
\Delta_1(\Lambda) &=-\left[1+\f{m_\pi^2}{m_K^2}\right]I(m_\pi^2)-
3\,\left[1+\f{m_8^2}{m_K^2}\right]I(m_8^2)-4 m_K^2 I^\prime(m_K^2)
\\
\Delta_5(\Lambda)& =+\left[1+\f{m_\pi^2}{m_K^2}\right]I(m_\pi^2)+
3\, \left[1+\f{m_8^2}{m_K^2}\right]I(m_8^2)-4 m_K^2 I^\prime(m_K^2)
\end{aligned}
\ee
and
\be\label{I2}
I(m_i^2)=\frac{i}{(2\pi)^4}\int\frac{d^4q}{q^2-m_i^2}=\frac{1}{16\pi^2}\left[\Lambda^2-m_i^2\ln(1+\frac{\Lambda^2}{m_i^2})\right]
\ee

\be\label{I3}
I^\prime(m_i^2)\equiv\frac{dI(m_i^2)}{dm_i^2}=\frac{1}{16\pi^2}\left[\frac{\Lambda^2}{\Lambda^2+m_i^2}-\ln(1+\frac{\Lambda^2}{m_i^2})\right].
\ee

Using the well-known Gell-Mann-Okubo mass relation
\be
m_\pi^2+3 m_8^2= 4 m_K^2
\ee
we obtain
\be
B_1(0.6\gev)=0.75\, (1-0.11)\,, \qquad B_1(0.7\gev)=0.75\ (1-0.18)
\ee
and
\be
B_5(0.6\gev)=0.23\, (1+0.19)\,, \qquad B_5(0.7\gev)=0.23\ (1+0.29)\,.
\ee

For $B_1$, the resulting DQCD evolution is quite impressive with
\be
B_1(0)=0.75\,\Rightarrow\, B_1(0.7\gev)=0.62~||~0.61= B_1(1\gev)\,\Leftarrow\, 0.53
=B_1(3\gev).
\ee
The values of $B_1$  at the end of meson and quark-gluon evolutions 
 are close to each other and are expected to be
equal when the gap $||$ between these two evolutions will be filled by 
vector mesons.
For $B_5$, the resulting DQCD evolution  is displayed in Table~\ref{tab:epe-LQ-contr}.

\renewcommand{\refname}{R\lowercase{eferences}}

\addcontentsline{toc}{section}{References}

\bibliographystyle{JHEP}
\bibliography{Bookallrefs}
\end{document}